\pdfoutput=1
\RequirePackage[T1]{fontenc}

\documentclass[a4paper,11pt]{article}
\usepackage{jcappub}

\usepackage{amsmath}
\usepackage{mathtools}
\usepackage{mathrsfs}
\usepackage{amssymb}
\usepackage{mathtools}
\usepackage{slashed}
\usepackage{physics,braket}
\usepackage{graphicx,rotating}
\usepackage{xcolor}
\usepackage[utf8]{inputenc}
\usepackage{siunitx}
\usepackage{acronym}
\usepackage{bm}
\usepackage{bbm}
\usepackage{times}
\usepackage{mdframed}
\usepackage{datetime}
\usepackage{dsfont}
\usepackage{multirow}
\usepackage{cancel}
\usepackage{pifont}
\usepackage{romannum}
\usepackage{chngcntr}
\usepackage[most]{tcolorbox}
\usepackage{leftindex}
\usepackage{booktabs}
\usepackage{longtable}
\usepackage{array}
\usepackage{adjustbox}

\definecolor{darkred}{rgb}{0.7, 0., 0.}
\definecolor{mblue}{HTML}{0019A2}
\definecolor{monza}{HTML}{CF000F}
\definecolor{darkmagenta}{HTML}{8b008b}
\definecolor{darkgreen}{HTML}{006E4F}
\definecolor{lightpink}{rgb}{1,0.4,0.4}
\definecolor{carmine}{rgb}{0.59, 0.0, 0.09}
\AtBeginDocument{\pagenumbering{arabic}}
\numberwithin{equation}{section}

\def\kc{k_\mathrm{c}}

\newcommand{\ee}{\mathrm{e}}

\newcommand{\calC}{\mathcal{C}}

\newcommand{\scrF}{\mathscr{F}}

\newcommand{\scrG}{\mathscr{G}}

\newcommand{\calL}{\mathcal{L}}
\newcommand{\calY}{\mathcal{Y}}

\newcommand{\calX}{\mathcal{X}}

\newcommand{\bmk}{\bm{k}}
\newcommand{\bmd}{\bm{d}}
\newcommand{\bmx}{\bm{x}}
\newcommand{\bmp}{\bm{x}_{\rm p}}
\newcommand{\hbmk}{\hat{\bm{k}}}
\newcommand{\hbmd}{\hat{\bm{d}}}
\newcommand{\hbmx}{\hat{\bm{x}}}
\newcommand{\hbmp}{\hat{\bm{x}}_{\rm p}}

\newcommand{\sY}[3]{\leftindex_{#1}{Y}_{#2 #3}}

\newcommand{\bae}[1]{\begin{align} #1 \end{align}}

\newcommand{\beae}[1]{\begin{equation}\begin{aligned} #1 \end{aligned}\end{equation}}

\newcommand{\hatk}{\hat{k}}
\newcommand{\hatx}{\hat{x}}

\newcommand{\multi}[1]{\begin{multline} #1 \end{multline}}

\newcommand{\pistd}{\hat{\pi}_{\mathrm{std}}}
\newcommand{\pimod}{\hat{\pi}_{\mathrm{mod}}}
\newcommand{\Pistd}{\hat{\Pi}_{\mathrm{std}}}
\newcommand{\Pimod}{\hat{\Pi}_{\mathrm{mod}}}

\newcommand{\Hquad}{\hspace{.25em}} 

\newcommand{\largelike}{\fontsize{14pt}{14pt}\selectfont}

\acrodef{EoM}{equation of motion}
\acrodef{CMB}{cosmic microwave background}
\acrodef{LSS}{large-scale structure}
\acrodef{LOS}{line-of-sight}
\acrodef{IA}{intrinsic alignment}
\acrodef{PP}{plane-parallel}
\acrodef{SFB}{sherical Fourier-Bessel}
\acrodef{BipoSH}{bipolar spherical harmonics}
\acrodef{TripoSH}{tripolar spherical harmonics}
\acrodef{RSD}{redshift-space distortion}
\acrodef{CG}{Clebsch--Gordan}

\makeatletter
\def\fnum@figure{\normalfont\figurename\nobreakspace\thefigure}
\def\fnum@table{\normalfont\tablename\nobreakspace\thetable}
\makeatother
\begin{document}

\begin{titlepage}
  \hfill\makebox[0pt][r]{KEK-TH-2857, KEK-Cosmo-0427}
  \vskip .5in
  
  \begin{center}
    {\Huge \bfseries
    Dipolar power asymmetry 
    \\[.1in]
   in wide-angle correlations of
    \\[.2in]
    galaxy density, velocity and ellipticity
    }	
  \end{center}
        \vskip .4in
   \begin{center}
   {\largelike
        Yusuke Mikura,$^{\ast}$\footnote{\href{mailto:ymikura@asiaa.sinica.edu.tw}{ymikura@asiaa.sinica.edu.tw}}
        \Hquad
        Teppei Okumura,$^{\ast \dag}$\footnote{\href{mailto:tokumura@asiaa.sinica.edu.tw}{tokumura@asiaa.sinica.edu.tw}}
        \Hquad
        Ippei Obata,$^{\$ \dag}$\footnote{\href{iobata@post.kek.jp}{iobata@post.kek.jp}}
        \Hquad
        Maresuke Shiraishi$^{\S}$\footnote{\href{mailto:shiraishi@onomichi-u.ac.jp}{shiraishi@onomichi-u.ac.jp}}
        }
  \end{center}
  \vskip .3in
      \begin{center}
        \def\arraystretch{1}
      \begin{tabular}{ll}
          $^\ast$& 
          Institute of Astronomy and Astrophysics
          \\
          & Academia Sinica, 
          \\
          & No. 1, Section 4, Roosevelt Road, Taipei 106319, Taiwan
          \\[0.5em]
          $^\dag$& 
          Kavli Institute for the Physics and Mathematics of the Universe (WPI), \\
          & UTIAS, The University of Tokyo, 
          \\
          & Kashiwa, Chiba 277-8583, Japan
          \\[0.5em]          
          $^\$ $& 
          Theory Center, Institute of Particle and Nuclear Studies (IPNS), 
          \\
          & High Energy Accelerator Research Organization (KEK),
          \\
          & 1-1 Oho, Tsukuba, Ibaraki 305-0801, Japan
          \\[0.5em]
          $^\S$& 
          Department of Economics, Management and Information Science, 
          \\
          & Onomichi City University, 
          \\
          & Onomichi, Hiroshima 722-8506, Japan
      \end{tabular}
    \end{center}
    \vskip .4in
    \noindent\ignorespaces
    {\bfseries Abstract:}
    The large-scale structure of the universe has the potential to probe anomalies suggested by observations of the cosmic microwave background. In this work, we focus on a position-dependent dipolar modulation of the primordial power spectrum and develop a full-sky formalism for computing correlation functions of galaxy density, velocity and ellipticity. By comparing the correlation functions obtained with and without the plane-parallel approximation, we show that wide-angle corrections become non-negligible for opening angles $\Theta \gtrsim 30^\circ$. Our results demonstrate that wide-angle corrections must be taken into account when testing the dipolar modulation with future large-scale structure surveys.
    \end{titlepage}

\setcounter{tocdepth}{2}
{
  \hypersetup{linkcolor=darkred}
  \beforetochook\hrule
  \tableofcontents
  \afterTocSpace\hrule
\pagestyle{myplain}\pagenumbering{arabic}
}
\flushbottom
\acresetall
\section{Introduction}
Inflation provides a compelling description of the primordial universe. The simplest inflationary scenario, single-field slow-roll inflation, predicts that cosmological fluctuations are nearly isotropic and homogeneous on large scales. While these statistical properties are supported by observations of the \ac{CMB}, several anomalies have been reported that may originate from physics beyond the simplest scenario~\cite{Schwarz:2015cma}. 

One example of these anomalies is a hemispherical power asymmetry or dipolar power modulation, where the amplitude of the temperature fluctuations in one hemisphere is larger than in the other. This power asymmetry may arise when the density perturbation has statistical inhomogeneity, modeled as~\cite{Gordon:2006ag} 
\bae{
\delta_{\rm m} (\bm{k}, \hbmx)  = \bar{\delta}_{\rm m} (\bm{k})\left(1 + A (k) \hbmd \cdot \hbmx \right) ~,
}
where $\bar{\delta}_{\rm m} (\bm{k})$ is the Fourier mode of the homogeneous and isotropic density fluctuation, $A (k)$ is the scale-dependent amplitude of the dipolar asymmetry, and $\hbmd \cdot \hbmx$ denotes the inner product between the preferred direction and the \ac{LOS}. The amplitude of the modulation on large scales has been reported to be $A (k) \simeq 0.07$ by the WMAP~\cite{Eriksen:2003db,Hansen:2004vq,Eriksen:2007pc,Hoftuft:2009rq} and Planck collaborations~\cite{Planck:2013lks,Planck:2015igc,Planck:2019evm} with a significance of $3\sigma$. Further analyses have included small-scale \ac{CMB} data, suggesting that the dipolar modulation exhibits a scale dependence~\cite{Duncan,Flender:2013jja}. In ref.~\cite{Aiola:2015rqa}, the authors considered a power-law model and found that the modulation amplitude scales as $A(k) \propto k^{-1/2}$. Theoretically, the dipolar modulation could be generated by, e.g., a linear gradient of a long-wavelength perturbation whose wavelength is much larger than the horizon scale at the last-scattering surface~\cite{Erickcek:2008sm,Erickcek:2008jp}. Nevertheless, despite these observational hints and theoretical developments~\cite{Erickcek:2009at,Schmidt:2012ky,Liddle:2013czu,Mazumdar:2013yta,Kanno:2013ohv,Lyth:2013vha,Kobayashi:2015qma,Ashoorioon:2015pia,Byrnes:2015dub,Byrnes:2016uqw}, the physical origin of the dipolar modulation remains unclear.

As a complement to the \ac{CMB}, the \ac{LSS} provides a powerful probe of possible violations of the statistical isotropy and homogeneity in the primordial fluctuations~\cite{Hirata:2009ar,Yoon:2014daa,Appleby:2014lra,Alonso:2014xca,Tiwari:2015tba,Bengaly:2016amk}. One key advantage of the \ac{LSS} over the \ac{CMB} is its access to three-dimensional information. Indeed, ref.~\cite{Shiraishi:2016wec} demonstrated that the three-dimensional distribution of galaxies can significantly improve constraints on the primordial anisotropy. 
Another advantage is the availability of multiple complementary observables, including galaxy distribution, peculiar velocity and galaxy ellipticity. While the galaxy distribution and peculiar velocity have been conventional observables of spin-weight $0$, it has been recently recognized that correlations of intrinsic spin-weight $2$ galaxy ellipticity, referred to as \ac{IA}, contain a wealth of cosmological information (see, e.g. refs.~\cite{Chisari:2013dda,Schmidt:2015xka,Chisari:2016xki,Kogai:2018nse,Okumura:2019ozd,Taruya:2020tdi,Kogai:2020vzz,Akitsu:2020jvx,Okumura:2021xgc,Saga:2022frj,Okumura:2023pxv,Kurita:2023qku,Okumura:2024xnd,Mikura:2025vaj,Kurita:2025hmp,Okumura:2026bpq}). 
The information provided by \ac{IA} can be used for testing possible  violation of the statistical isotropy. In ref.~\cite{Minato:2025ozy}, the authors combined galaxy clustering with \ac{IA} to examine the feasibility of testing the statistical anisotropy induced by the dipole modulation under the assumption of a small opening angle, namely, the \ac{PP} approximation.\footnote{The authors of ref.~\cite{Minato:2025ozy} recently extended their analysis by incorporating the peculiar velocity in ref.~\cite{Minato:2026ykj}.}

As forthcoming \ac{LSS} surveys will probe wide angular separations, it is essential to develop a comprehensive formalism beyond the \ac{PP} approximation. Indeed, neglecting wide-angle effects in the correlation functions involving the galaxy number density, peculiar velocity and ellipticity can lead to errors exceeding $10\%$ at opening angles of several tens of degrees~\cite{Szapudi:2004gh,Yoo:2013zga,Castorina:2017inr,Taruya:2019xsf,Castorina:2019hyr,Shiraishi:2020vvj,Shiraishi:2021oau,Shiraishi:2023zda}. Under the \ac{PP} approximation, two \ac{LOS} directions toward galaxies located at $\bm{x}_1$ and $\bm{x}_2$ are approximated by a single common vector $\bmp$. By allowing for the possibility of both zero and nonzero spin weights, the correlation functions in the \ac{PP} limit can be decomposed by the spin-weighted \ac{BipoSH}, $\{\bm{Y}_{\ell}(\hbmx_{12}) \otimes \leftindex_{\lambda}{\bm{Y}}_{\ell^\prime}(\hbmp)\}_{LM}$, where $\bmx_{12} \coloneqq \bm{x}_1 - \bm{x}_2$~\cite{Shiraishi:2020vvj,Shiraishi:2023zda}. For galaxy pairs with a large opening angle, however, the two \ac{LOS} directions must be treated separately. Therefore, the correlation functions depend on two vectors $\bmx_1$ and $\bmx_2$. In this situation, the spin-weighted \ac{TripoSH}, $\{\bm{Y}_{\ell}(\hbmx_{12}) \otimes \{\leftindex_{\lambda_1}{\bm{Y}}_{\ell_1}(\hbmx_1) \otimes \leftindex_{\lambda_2}{\bm{Y}}_{\ell_2}(\hbmx_2)\}{}_{\ell^\prime} \}_{LM}$, form a natural basis for expanding the correlation functions in configuration space~\cite{Shiraishi:2020vvj,Shiraishi:2023zda}.
These decomposition techniques based on the spin-weighted polypolar spherical harmonics have been used as a powerful tool for computing galaxy statistics. They have been applied to wide-angle effects of isotropic spin-weight $0$ fields~\cite{Szalay:1997cc,Szapudi:2004gh,Yoo:2013zga,Castorina:2017inr,Taruya:2019xsf,Castorina:2019hyr,Shiraishi:2021oau} and spin-weight $2$ field~\cite{Shiraishi:2020vvj}, as well as to isotropy-breaking signatures from fields with spin-weight $0$~\cite{Shiraishi:2016wec,Bartolo:2017sbu,Akitsu:2019avy,Shiraishi:2020pea} and fields with finite spin-weight~\cite{Shiraishi:2023zda}.\footnote{A modulation pattern discussed in ref.~\cite{Shiraishi:2023zda} is different from the position-dependent one.} As for the dipolar modulation, ref.~\cite{Shiraishi:2016wec} investigated the \ac{TripoSH} decomposition for the wide-angle two-point correlation function of  the galaxy density field. 

The aim of this paper is to provide a theoretical foundation for testing the dipolar modulation with three-dimensional two-point correlation functions in future wide-angle \ac{LSS} surveys. In this work, building on the formalism developed in ref.~\cite{Shiraishi:2023zda}, we generalize ref.~\cite{Shiraishi:2016wec} by including the Doppler term in the density field and by incorporating the radial component of peculiar velocity and the ellipticity. In particular, we compute the two-point correlation functions of the galaxy observables using the spin-weighted \ac{TripoSH} and show that the full-sky treatment is important for an accurate test of the dipolar asymmetry.

This paper is organized as follows. In section~\ref{sec. Galaxy observables}, we present redshift-space expressions for the galaxy density, velocity and ellipticity fields in the presence of the dipolar modulation. In section~\ref{sec. Computation methodology}, we describe an efficient computational method for three-dimensional correlation functions using the spin-weighted \ac{TripoSH}. We also derive the corresponding expressions in the \ac{PP} limit. In section~\ref{sec. Numerical computation}, we demonstrate the need for the full-sky treatment. Finally, we conclude in section~\ref{sec. Conclusions}. Appendix~\ref{sec. mathematics} provides a brief summary of mathematical definitions and identities used in this work. We derive a unified expression of the galaxy observables in terms of the spin-weighted spherical harmonics in appendix~\ref{sec. derivation unified}. We check the consistency of our results with known results in the \ac{PP} limit in appendix~\ref{sec. check of PP limit}.
\section{Galaxy observables in the presence of dipolar modulation}\label{sec. Galaxy observables}
In this section, following ref.~\cite{Shiraishi:2023zda}, we provide expressions for the galaxy density, velocity and ellipticity fields in redshift space using the spin-weighted spherical harmonics $\leftindex_{\lambda}{Y}_{\ell m}$.
\subsection{Matter fluctuation with a dipolar modulation}
Motivated by the dipolar modulation introduced in the context of the \ac{CMB}, we consider a small dipolar modulation of the matter density fluctuation. Without assuming a specific form for the modulation, we parametrize it as~\cite{Shiraishi:2016wec}\footnote{In this work, we focus on the position-dependent modulation. There is another type of modulation in which the power spectrum $P(\bmk)$ depends not only on the magnitude of the wave vector $\bmk$ but also on its direction. We refer to this as $g_{LM}$-type modulation. It may arise in the context of anisotropic inflation models~\cite{Watanabe:2009ct,Watanabe:2010fh,Kanno:2010nr,Watanabe:2010bu,Soda:2012zm,Bartolo:2012sd,Ohashi:2013qba,Ohashi:2013pca,Bartolo:2014hwa,Naruko:2014bxa,Bartolo:2015dga,Ito:2015sxj,Abolhasani:2015cve,Ito:2017bnn,Bartolo:2017sbu,Fujita:2018zbr}. See also refs.~\cite{Shiraishi:2016wec,Bartolo:2017sbu,Sugiyama:2017ggb,Shiraishi:2020pea,Shiraishi:2023zda} for observational aspects related to the \ac{LSS}.}
\bae{\label{eq. density spherical expression}
\delta_{\rm m} (\bm{k},  \hbmx) = \bar{\delta}_{\rm m} (\bm{k})\left[1 + \sum_{M} A_{1M}(k) Y_{1 M} (\hbmx)\right] ~,
}
where $\delta_{\rm m} (\bm{k}, \hbmx)$ is the local Fourier transformation, defined through
\bae{
\delta_{\rm m} (\bm{x}) = \int\frac{\dd[3]{k}}{(2\pi)^3}\ee^{i \bm{k}\cdot\bm{x}} \delta_{\rm m} (\bm{k}, \hbmx) ~.
}
Here, $\bar{\delta}_{\rm m} (\bm{k})$ is the homogeneous and isotropic part of the density fluctuation. The reality of the density field imposes $A^\ast_{1M} = (-1)^M A_{1 -M}$. Assuming $|A_{1M}|\ll1$, the two-point correlation function of the matter density field is defined by
\bae{
  \braket{\delta_{\rm m} (\bm{x}_1) \delta_{\rm m} (\bm{x}_2)}  = 
  \int \frac{\dd^3 k}{(2\pi)^3}
  \ee^{i\bm{k}\cdot (\bm{x}_1 - \bm{x}_2)} P_{\rm m} (\bmk, \hbmx_1, \hbmx_2) ~,
}
where
\bae{
  P_{\rm m} (\bmk, \hbmx_1, \hbmx_2) \simeq \bar{P}_{\rm m} (k) \left[1 + \sum_{M} A_{1M}(k) \left\{Y_{1 M} (\hbmx_1) + Y_{1 M} (\hbmx_2)\right\}\right] ~.
}
Here, $\bar{P}_{\rm m}(k)$ is the isotropic matter power spectrum, defined by
\bae{
  \braket{\bar{\delta}_{\rm m} (\bm{k}) \bar{\delta}_{\rm m} (\bm{k}^\prime)} 
  =
  \left(2\pi\right)^3\delta^{(3)}_{D}(\bm{k}+\bm{k}^\prime) \bar{P}_{\rm m} (k) ~,
}
where $\delta^{(3)}_D$ is the three-dimensional Dirac delta function.
\subsection{Galaxy density, velocity and ellipticity}
As conventional galaxy observables, let us first consider fields with spin-weight $0$: the galaxy density and peculiar velocity. The relation between the observed galaxy distribution in redshift space and the underlying matter distribution is given by the Kaiser formula~\cite{Kaiser:1987qv, Hamilton:1997zq, Yoo:2013zga}
\bae{\label{eq. Number density field}
\delta (\bm{x}) = 
\int\frac{\dd[3]{k}}{(2\pi)^3}
\ee^{i \bm{k}\cdot\bm{x}}
\left[b_g - i\frac{\alpha (\bm{x}) f}{kx}(\hbmk\cdot\hbmx)+ f (\hbmk\cdot\hbmx)^2\right] \delta_{\rm m} (\bm{k}, \hbmx) ~,
}
where $b_g$ is the linear bias, $f$ is the linear growth rate, and $\alpha (\bm{x})$ is the selection function of a given galaxy sample. The second term is called the Doppler term~\cite{Bonvin:2011bg,Raccanelli:2016avd} which should not be neglected in future wide-angle surveys. If one assumes a uniform radial selection function, we have $\alpha (\bm{x}) =2$~\cite{Yoo:2013zga}. 
In contrast to the galaxy density, the peculiar velocity can be treated as bias-free on large scales, assuming that galaxies and matter move along the same trajectories. Then, the \ac{LOS} component of the peculiar velocity is simply given by~\cite{Gorski,Hamilton:1997zq}
\bae{
\label{eq. peculiar velocity}
u (\bm{x}) = 
\int\frac{\dd[3]{k}}{(2\pi)^3}
\ee^{i \bmk\cdot \bmx} 
i \frac{a H f}{k}(\hbmk\cdot\hbmx)\delta_{\rm m} (\bm{k}, \hbmx) ~,
}
where $a$ is the scale factor and $H$ is the Hubble parameter.

In the linear alignment model~\cite{Catelan:2000vm,Hirata:2004gc,Blazek:2011xq,Okumura:2019ozd}, the intrinsic ellipticity is modeled as a symmetric and trace-free projection of the matter tidal field on the plane orthogonal to the \ac{LOS}:
\bae{
  \gamma_{ij} (\bm{x}) = P_{ij}{}^{k\ell} (\hbmx)\int\frac{\dd[3]{k}}{(2\pi)^3}\ee^{i \bm{k}\cdot\bm{x}}\left(\hatk_k \hatk_\ell - \frac13\delta_{k\ell}\right) b_{\rm K} \delta_{\rm m} (\bm{k}, \hbmx) ~,
}
where $b_{\rm K}$ is a linear shape bias parameter and $P_{ij}{}^{k\ell}(\hbmx)$ is a projector onto the plane orthogonal to the \ac{LOS} direction $\hbmx$, defined by
\bae{
  P_{ij}{}^{k\ell}(\hbmx) = q_{i}{}^k(\hbmx) q_{j}{}^\ell (\hbmx) - \frac12 q_{ij}(\hbmx) q^{k\ell}(\hbmx) ~.
  }
Here, $q_{ij}(\hbmx)$ is the induced metric on the projected plane
\bae{
  q_{ij}(\hbmx) = g_{ij} - \hatx_i \hatx_j ~, \quad \mathrm{with} \quad \hatx_i \hatx^i = 1 ~.
}
Since the ellipticity field is defined on a two-sphere $S^2$, it is convenient to introduce functions of spin-weight $\pm2$. 
Let $\bm{e}_1$ and $\bm{e}_2$ be two orthogonal vectors, that are associated to the \ac{LOS} direction $\hbmx$. We can then define basis vectors by
\bae{
  \bm{m}_{\pm} (\hbmx) \coloneqq \frac{1}{\sqrt{2}}\left(\bm{e}_1 \mp i \bm{e}_2 \right) ~,
}
where the set $\{\bm{m}_{+}, \bm{m}_{-}, \hat{\bm{x}}\}$ forms a new frame satisfying
\bae{\label{eq. m normality}
\bm{m}_{\pm}\cdot\bm{m}_{\pm} = 0 ~,
\quad 
\bm{m}_{\pm}\cdot\bm{m}_{\mp} = 1 ~,
\quad 
\bm{m}_{\pm}\cdot\hat{\bm{x}} = 0 ~.
}
Using the basis vectors $\bm{m}_{\pm}$, the spin-weighted functions for the ellipticity field are defined by
\bae{\label{eq. spin-weight ellipticity}
  \leftindex_{\pm 2}{\gamma} (\bm{x}) \coloneqq m_{\mp}^i (\hbmx) m_{\mp}^j (\hbmx) \gamma_{ij} (\bm{x}) 
  = \int\frac{\dd[3]{k}}{(2\pi)^3}\ee^{i \bm{k}\cdot\bm{x}} m_{\mp}^i (\hbmx) m_{\mp}^j (\hbmx) \hatk_i \hatk_j b_{\rm K} \delta_{\rm m} (\bm{k}, \hbmx) ~.
}
\subsection{Unified expression}
For later convenience, we provide a unified expression for the three fields: density $\delta$, velocity $u$, and ellipticity $\leftindex_{\pm 2}{\gamma}$, in terms of the spin-weighted spherical harmonics $\leftindex_{\lambda}{Y}_{\ell m}$. The expressions for those fields can be cast into~\cite{Shiraishi:2020vvj,Shiraishi:2023zda}
\bae{\label{eq. unified expression}
\leftindex_{\lambda}{X} (\bm{x}) = \int\frac{\dd[3]{k}}{(2\pi)^3}\ee^{i \bm{k}\cdot\bm{x}}
  \sum_{j\mu} \frac{4\pi}{2j+1} c_j^{X} Y_{j\mu} (\hbmk)\leftindex_{-\lambda}{Y}^\ast_{j\mu} (\hbmx) 
  \delta_{\rm m} (\bm{k}, \hbmx) ~,
}
where $X \in \{\delta, u, \gamma\}$ and $c_j^{X}$ are defined by (see appendix~\ref{sec. derivation unified} for the derivation)
\bae{
  c_j^\delta & \coloneqq \left[\left(b_g+\frac{1}{3}f\right) \delta^{\rm K}_{j,0} - i\frac{\alpha(\bm{x}) f}{kx}\delta^{\rm K}_{j,1}+\frac{2}{3}f\delta^{\rm K}_{j,2} \right] \delta^{\rm K}_{\lambda,0}  ~,
  \\
  c_j^u & \coloneqq i \frac{a H f}{k} \delta^{\rm K}_{j,1} \delta^{\rm K}_{\lambda, 0} ~,
  \\
  c_j^{\gamma} & \coloneqq \frac{\sqrt{6}}{3} b_{\rm K} \delta^{\rm K}_{j,2} ~.
}
Here, $\delta^{\rm K}_{i,j}$ is the Kronecker delta.
For notational brevity, we omit the spin-weight label $0$ for the density and velocity fields.
\section{Computation methodology}\label{sec. Computation methodology}
This section provides a theoretical foundation to calculate the two-point correlation functions in the presence of the dipolar modulation. In section~\ref{sec. Correlation function and position-dependent power spectrum}, we define the correlation function and a related quantity. 
We then develop efficient computational methods for full-sky correlations in section~\ref{sec. Spin-weighted TripoSH expansion} and for the \ac{PP} limit in section~\ref{sec. The plane-parallel limit}.
\subsection{Correlation function and position-dependent power spectrum}\label{sec. Correlation function and position-dependent power spectrum}
We define the correlation function of fields $X_1, X_2 \in \{\delta, u, \gamma\}$ by 
\bae{
  \xi_{\lambda_1 \lambda_2}^{X_1 X_2} (\bm{x}_{12}, \hbmx_1, \hbmx_2) \coloneqq \braket{\leftindex_{\lambda_1}{X}_{1} (\bm{x}_1)\leftindex_{\lambda_2}{X}_{2} (\bm{x}_2)} 
  =
  \int\frac{\dd[3]{k}}{(2\pi)^3}\ee^{i \bm{k}\cdot\bm{x}_{12}} P_{\lambda_1 \lambda_2}^{X_1 X_2} (\bm{k}, \hbmx_1, \hbmx_2) ~,
}
where $\bm{x}_{12}\coloneqq \bm{x}_{1}-\bm{x}_{2}$. We note that $P_{\lambda_1 \lambda_2}^{X_1 X_2}$ is not an exact Fourier counterpart of the correlation function $\xi_{\lambda_1 \lambda_2}^{X_1 X_2}$ because the position dependence remains in its expression. In this paper, we call it position-dependent power spectrum. Using the unified expression~\eqref{eq. unified expression}, one finds that the position-dependent power spectrum takes the form
\multi{\label{eq. Power-like}
P_{\lambda_1 \lambda_2}^{X_1 X_2} (\bm{k}, \hbmx_1, \hbmx_2)
= \sum_{j_1\mu_1 j_2\mu_2}
\frac{16\pi^2 (-1)^{j_2} c_{j_1}^{X_1}c_{j_2}^{X_2}}{(2j_1+1)(2j_2+1)}
Y_{j_1\mu_1}(\hbmk)Y_{j_2\mu_2}(\hbmk)
\leftindex_{-\lambda_1}{Y}^\ast_{j_1\mu_1} (\hbmx_1) \leftindex_{-\lambda_2}{Y}^\ast_{j_2\mu_2} (\hbmx_2) 
\\
\times \left[1 + \sum_{M} A_{1 M}(k) \left\{Y_{1 M} (\hbmx_1) + Y_{1 M} (\hbmx_2)\right\}\right] \bar{P}_{\rm m} (k) ~,
}
where we used the parity transformation of the spin-weighted spherical harmonics
\bae{
\leftindex_{\lambda}{Y}_{\ell m} (-\hbmx) = (-1)^\ell \leftindex_{-\lambda}{Y}_{\ell m} (\hbmx) ~.
}
Using properties of the spin-weighted spherical harmonics listed in appendix~\ref{sec. The spin-weighted spherical harmonics}, one can simplify the form of the position-dependent power spectrum as
\bae{\label{eq. Power-like simplified}
P_{\lambda_1 \lambda_2}^{X_1 X_2} (\bm{k}, \hbmx_1, \hbmx_2) = \sum_{j_1\mu_1 j_2\mu_2 J \mu}
\mathcal{K}^{J\mu}_{j_1\mu_1 j_2\mu_2}
c_{j_1}^{X_1} c_{j_2}^{X_2}
Y_{J-\mu}(\hbmk)
\left[\calY^{(0)} + \calY^{(1)} + \calY^{(2)}\right] \bar{P}_{\rm m}(k) ~,
}
where
\bae{
\mathcal{K}^{J\mu}_{j_1\mu_1 j_2\mu_2} 
& \coloneqq (-1)^{\lambda_1 + \lambda_2 + j_2 + \mu}\sqrt{\frac{64\pi^3}{(2j_1+1)(2j_2+1)(2J+1)}} \calC^{J0}_{j_1 0 j_2 0} \calC^{J-\mu}_{j_1 \mu_1 j_2 \mu_2} ~,
}
and 
\bae{
\calY^{(0)}
& \coloneqq \leftindex_{\lambda_1}{Y}_{j_1 -\mu_1}(\hbmx_1) \leftindex_{\lambda_2}{Y}_{j_2 -\mu_2}(\hbmx_2) ~,
\\
\calY^{(1)}
& \coloneqq \sum_{j^\prime \rho M}
\sqrt{\frac{3 (2 j_1+1)}{4\pi (2j^\prime+1)}}
\calC^{j^\prime -\lambda_1}_{j_1 -\lambda_1 1 0}
\calC^{j^\prime-\rho}_{j_1 -\mu_1 1 M}
\leftindex_{\lambda_1}{Y}_{j^\prime -\rho}(\hbmx_1) \leftindex_{\lambda_2}{Y}_{j_2 -\mu_2}(\hbmx_2) A_{1M}(k) ~,
\\
\calY^{(2)}
& \coloneqq \sum_{j^\prime \rho M}
\sqrt{\frac{3 (2 j_2+1)}{4\pi (2j^\prime+1)}}
\calC^{j^\prime -\lambda_2}_{j_2 -\lambda_2 1 0}
\calC^{j^\prime-\rho}_{j_2 -\mu_2 1 M} 
\leftindex_{\lambda_1}{Y}_{j_1 -\mu_1}(\hbmx_1)
\leftindex_{\lambda_2}{Y}_{j^\prime -\rho}(\hbmx_2) A_{1M}(k) ~.
}
Here, $\calC^{LM}_{\ell_1 m_1 \ell_2 m_2}$ are the \ac{CG} coefficients in the Condon--Shortley phase convention, which are written in terms of the Wigner $3j$ symbol as~\cite{Varshalovich:1988ifq}
\bae{
 \calC^{LM}_{\ell_1 m_1 \ell_2 m_2} = (-1)^{\ell_1-\ell_2+M}\sqrt{2L+1}
 \left(\begin{matrix}
 \ell_1 & \ell_2  & L\\
     m_1 & m_2 & -M
 \end{matrix}  \right) ~.
}
The \ac{CG} coefficients vanish unless the following conditions are satisfied:
\bae{\label{eq. triangle}
|\ell_1-\ell_2| \leq L \leq \ell_1 +\ell_2 ~,
\\
m_1 + m_2 = M ~,
}
where $|m_1|\leq \ell_1$, $|m_2| \leq \ell_2$ and $|M| \leq L $.
We denote the triangle condition~\eqref{eq. triangle} by $\triangle (\ell_1, \ell_2, L)$, following ref.~\cite{Graphical_theory}. 
\subsection{Spin-weighted TripoSH expansion}\label{sec. Spin-weighted TripoSH expansion}
When discussing two-point correlations of the spin-weighted functions in three dimensions, the spin-weighted \ac{TripoSH} serve as a natural basis for the decomposition, since the correlation functions depend on three directions, $\hbmx_1$, $\hbmx_2$, and $\hbmx_{12}$, as well as on two spin-weight indices, $\lambda_1$ and $\lambda_2$~\cite{Shiraishi:2020vvj,Shiraishi:2023zda}. Let $\leftindex_{\lambda}{\bm{Y}}_{\ell}$ denote the $(2\ell+1)$-component multiplet formed by the spin-weighted spherical harmonics. The spin-weighted \ac{TripoSH} are then defined by~\cite{Shiraishi:2023zda} 
\beae{\label{eq. TripoSH basis}
    \leftindex_{\lambda_1 \lambda_2}{\calX}_{\ell\ell_1 \ell_2 \ell^\prime}^{LM}(\hbmx_{12}, \hbmx_{1}, \hbmx_{2}) 
    \coloneqq ~&
    \{\bm{Y}_{\ell}(\hbmx_{12}) \otimes 
    \{\leftindex_{\lambda_1}{\bm{Y}}_{\ell_1}(\hbmx_1) \otimes \leftindex_{\lambda_2}{\bm{Y}}_{\ell_2}(\hbmx_2)\}{}_{\ell^\prime} \}_{LM} ~,
    \\ 
    = ~& \sum_{m m^\prime m_1 m_2} 
   \calC^{LM}_{\ell m \ell^\prime m^\prime} \calC^{\ell^\prime m^\prime}_{\ell_1 m_1 \ell_2 m_2} 
   Y_{\ell m}(\hbmx_{12}) 
   \sY{\lambda_1}{\ell_1}{m_1}(\hbmx_1) \sY{\lambda_2}{\ell_2}{m_2} (\hbmx_2) ~.
}
With these, we can decompose the correlation function as 
\bae{\label{eq. hatXi}
\xi_{\lambda_1 \lambda_2}^{X_1 X_2} (\bm{x}_{12}, \hbmx_1, \hbmx_2) = \sum_{\ell\ell_1 \ell_2 \ell^\prime L M}
  \leftindex_{\lambda_1 \lambda_2}{\left(\hat{\Xi}^{X_1 X_2}\right)}_{\ell\ell_1 \ell_2 \ell^\prime}^{LM} (x_{12})
   \leftindex_{\lambda_1 \lambda_2}{\calX}_{\ell\ell_1 \ell_2 \ell^\prime}^{LM}(\hbmx_{12}, \hbmx_{1}, \hbmx_{2})  ~,
}
and, similarly, the position-dependent power spectrum as
\bae{\label{eq. hatPi}
P_{\lambda_1 \lambda_2}^{X_1 X_2} (\bm{k}, \hbmx_1, \hbmx_2) = \sum_{\ell\ell_1 \ell_2 \ell^\prime L M}
  \leftindex_{\lambda_1 \lambda_2}{\left(\hat{\Pi}^{X_1 X_2}\right)}_{\ell\ell_1 \ell_2 \ell^\prime}^{LM} (k)
   \leftindex_{\lambda_1 \lambda_2}{\calX}_{\ell\ell_1 \ell_2 \ell^\prime}^{LM}(\hbmk, \hbmx_{1}, \hbmx_{2}) ~.
}
Here and below, we denote expansion coefficients with a hat.
One can show that the coefficients $\hat{\Xi}$ and $\hat{\Pi}$ are related through a Hankel transform as
\bae{
 \leftindex_{\lambda_1 \lambda_2}{\left(\hat{\Xi}^{X_1 X_2}\right)}_{\ell\ell_1 \ell_2 \ell^\prime}^{LM} (x_{12}) = 
 i^\ell \int^\infty_0 \frac{k^2 \dd k}{2\pi^2} j_\ell (kx_{12}) \leftindex_{\lambda_1 \lambda_2}{\left(\hat{\Pi}^{X_1 X_2}\right)}_{\ell\ell_1 \ell_2 \ell^\prime}^{LM} (k) ~.
}
The coefficient $\hat{\Pi}$ is given by the following angular integral of the position-dependent power spectrum multiplied by the complex conjugate of the spin-weighted \ac{TripoSH}~\eqref{eq. TripoSH basis}:
\bae{
\leftindex_{\lambda_1 \lambda_2}{\left(\hat{\Pi}^{X_1 X_2}\right)}_{\ell\ell_1 \ell_2 \ell^\prime}^{LM} (k) = \int \dd\Omega_k \dd\Omega_{x_1}\dd\Omega_{x_2}
P_{\lambda_1 \lambda_2}^{X_1 X_2} (\bm{k}, \hbmx_1, \hbmx_2)
\leftindex_{\lambda_1 \lambda_2}{\calX}_{\ell\ell_1 \ell_2 \ell^\prime}^{LM\ast}(\hbmk, \hbmx_{1}, \hbmx_{2}) ~.
}
Here, we use the orthonormality of the spin-weighted \ac{TripoSH}, given by
\bae{\label{eq. orthonormality BipoSH}
\int \dd\Omega_k \dd\Omega_{x_1}\dd\Omega_{x_2}
\leftindex_{\lambda_1 \lambda_2}{\calX}_{\ell\ell_1 \ell_2 \ell^\prime}^{LM}(\hbmk, \hbmx_{1}, \hbmx_{2}) \leftindex_{\lambda_1 \lambda_2}{\calX}_{\widetilde{\ell}\widetilde{\ell}_1 \widetilde{\ell}_2 \widetilde{\ell}^\prime}^{\widetilde{L}\widetilde{M}\ast}(\hbmk, \hbmx_{1}, \hbmx_{2}) 
= 
\delta^{\rm K}_{L, \widetilde{L}}
\delta^{\rm K}_{M, \widetilde{M}}
\delta^{\rm K}_{\ell, \widetilde{\ell}} 
\delta^{\rm K}_{\ell_1, \widetilde{\ell}_1} 
\delta^{\rm K}_{\ell_2, \widetilde{\ell}_2}
\delta^{\rm K}_{\ell^\prime, \widetilde{\ell}^\prime} ~.
}

Let us now compute the coefficient $\hat{\Pi}$ using eq.~\eqref{eq. Power-like simplified} and properties of the spin-weighted spherical harmonics presented in appendix~\ref{sec. The spin-weighted spherical harmonics}. To this end, we write the coefficient $\hat{\Pi}$ as the sum of two contributions:
\bae{\label{eq. Pi full}
  \leftindex_{\lambda_1 \lambda_2}{\left(\hat{\Pi}^{X_1 X_2}\right)}_{\ell\ell_1 \ell_2 \ell^\prime}^{LM} (k)
  = 
  \leftindex_{\lambda_1 \lambda_2}{\left(\Pistd^{X_1 X_2}\right)}_{\ell\ell_1 \ell_2 \ell^\prime}^{LM} (k)
  +
  \leftindex_{\lambda_1 \lambda_2}{\left(\Pimod^{X_1 X_2}\right)}_{\ell\ell_1 \ell_2 \ell^\prime}^{LM} (k) ~,
}
where $\Pistd$ denotes the standard contribution in the statistically isotropic and homogeneous case, while $\Pimod$ represents the additional contribution induced by the dipolar modulation.
The standard contribution $\Pistd$ is non-vanishing when $L=M=0$ and takes a simple form
\bae{\label{eq. Pi iso}
  \leftindex_{\lambda_1 \lambda_2}{\left(\Pistd^{X_1 X_2}\right)}_{\ell\ell_1 \ell_2 \ell^\prime}^{LM} (k)
  = 
  (-1)^{\ell_1 +\lambda_1 +\lambda_2}\sqrt{\frac{64\pi^3}{(2 \ell_1+1)(2 \ell_2+1)}} \calC^{\ell 0}_{\ell_1 0\ell_2 0} c^{X_1}_{\ell_1} c^{X_2}_{\ell_2} \delta^{\rm K}_{L,0} \delta^{\rm K}_{M,0} \delta^{\rm K}_{\ell, \ell^\prime} \bar{P}_{\rm m} (k) ~,
}
which recovers the result in ref.~\cite{Shiraishi:2020vvj}. By contrast, the dipolar modulation has non-vanishing contribution for $L=1$, which can be seen from
\bae{\label{eq. Pi mod}
\leftindex_{\lambda_1 \lambda_2}{\left(\Pimod^{X_1 X_2}\right)}_{\ell\ell_1 \ell_2 \ell^\prime}^{LM} (k)
=
(-1)^{\lambda_1 + \lambda_2 + \ell_2  + \ell^\prime}\left[\scrF^{(1,2)} + (-1)^{\ell  + \ell^\prime}\scrF^{(2,1)} \right] A_{1 M}(k) \delta^{\rm K}_{L,1}\bar{P}_{\rm m} (k) ~,
}
where we have introduced
\bae{
\scrF^{(1,2)} & \coloneqq 4\pi \sum_j
\sqrt{\frac{(2\ell^\prime+1)(2\ell_1+1)}{(2j+1)(2\ell_2+1)}}
c_{j}^{X_1} c_{\ell_2}^{X_2}
\calC^{j \lambda_1}_{1 0 \ell_1 \lambda_1}
\calC^{\ell 0}_{j 0 \ell_2 0}
\begin{Bmatrix}
    \ell_1 & \ell_2 & \ell^\prime \\
    \ell & 1 & j
\end{Bmatrix} ~,
  \\
\scrF^{(2,1)} & \coloneqq \left. \scrF^{(1,2)}\right|_{(X_1, \ell_1, \lambda_1)\leftrightarrow (X_2, \ell_2, \lambda_2)} ~.
}
Here, $\begin{Bsmallmatrix}
    \ell_1 & \ell_2 & \ell_3 \\
    \ell_4 & \ell_5 & \ell_6
  \end{Bsmallmatrix}$ is the Wigner $6j$ symbol, which is invariant under any permutation of its columns or under swapping the upper and lower entries in any two columns. The $6j$ symbol is nonzero if the four triangle conditions $\triangle (\ell_1, \ell_2, \ell_3)$, $\triangle (\ell_1, \ell_5, \ell_6)$, $\triangle (\ell_4, \ell_2, \ell_6)$ and $\triangle (\ell_4, \ell_5, \ell_3)$ are satisfied. For $X_1=X_2=\delta$, eq.~\eqref{eq. Pi mod} reproduces the result in ref.~\cite{Shiraishi:2016wec}, up to the additional Doppler term.
\begin{table}[htbp]
\centering
\small
\renewcommand{\arraystretch}{1.2}
\begin{tabular}{|c|c|} 
\hline
$\Xi$ & $(\ell, \ell_1, \ell_2, \ell^\prime)$ 
\\ \hline
$\delta\delta$ &
\begin{tabular}[c]{@{}l@{}}
$(0, 0, 1, 1)$, $(0, 1, 0, 1)$, $(0, 1, 2, 1)$, $(0, 2, 1, 1)$, $(0, 2, 3, 1)$, $(0, 3, 2, 1)$, $(1, 0, 2, 2)$, $(1, 1, 1, 1)$, 
\\
$(1, 1, 3, 2)$, $(1, 2, 0, 2)$, $(1, 2, 2, 1)$, $(1, 3, 1, 2)$, $(2, 0, 1, 1)$, $(2, 0, 3, 3)$, $(2, 1, 0, 1)$, $(2, 1, 2, 1)$,
\\
 $(2, 1, 2, 2)$, $(2, 1, 2, 3)$, $(2, 2, 1, 1)$, $(2, 2, 1, 2)$, $(2, 2, 1, 3)$, $(2, 2, 3, 1)$, $(2, 2, 3, 2)$, $(2, 2, 3, 3)$, 
\\
$(2, 3, 0, 3)$, $(2, 3, 2, 1)$, $(2, 3, 2, 2)$, $(2, 3, 2, 3)$, $(3, 0, 2, 2)$, $(3, 1, 3, 2)$, $(3, 1, 3, 3)$, $(3, 1, 3, 4)$, 
\\
$(3, 2, 0, 2)$, $(3, 2, 2, 3)$, $(3, 3, 1, 2)$, $(3, 3, 1, 3)$, $(3, 3, 1, 4)$, $(4, 1, 2, 3)$, $(4, 2, 1, 3)$, $(4, 2, 3, 3)$, 
\\
$(4, 2, 3, 4)$, $(4, 2, 3, 5)$, $(4, 3, 2, 3)$, $(4, 3, 2, 4)$, $(4, 3, 2, 5)$ 
\end{tabular}
\\ \hline
$\delta u$ &
\begin{tabular}[c]{@{}l@{}}
$(0, 0, 1, 1)$, $(0, 1, 0, 1)$, $(0, 1, 2, 1)$, $(0, 2, 1, 1)$, $(1, 0, 0, 0)$, $(1, 0, 2, 2)$, $(1, 1, 1, 0)$, $(1, 1, 1, 1)$, 
\\
$(1, 1, 1, 2)$, $(1, 2, 0, 2)$, $(1, 2, 2, 0)$, $(1, 2, 2, 1)$, $(1, 2, 2, 2)$, $(1, 3, 1, 2)$, $(2, 0, 1, 1)$, $(2, 1, 0, 1)$, 
\\
$(2, 1, 2, 1)$, $(2, 1, 2, 2)$, $(2, 1, 2, 3)$, $(2, 2, 1, 1)$, $(2, 2, 1, 2)$, $(2, 2, 1, 3)$, $(3, 1, 1, 2)$, $(3, 2, 0, 2)$, 
\\
$(3, 2, 2, 2)$, $(3, 2, 2, 3)$, $(3, 2, 2, 4)$, $(3, 3, 1, 2)$, $(3, 3, 1, 3)$, $(3, 3, 1, 4)$ 
\end{tabular}
\\ \hline
$u u$ &
\begin{tabular}[c]{@{}l@{}}
$(0, 0, 1, 1)$, $(0, 1, 0, 1)$, $(0, 1, 2, 1)$, $(0, 2, 1, 1)$, $(2, 0, 1, 1)$, $(2, 1, 0, 1)$, $(2, 1, 2, 1)$, $(2, 1, 2, 2)$, 
\\
$(2, 1, 2, 3)$, $(2, 2, 1, 1)$, $(2, 2, 1, 2)$, $(2, 2, 1, 3)$ 
\end{tabular}
\\ \hline
$\leftindex_{\pm2}{\gamma}\leftindex_{\pm2}{\gamma}$ &
\begin{tabular}[c]{@{}l@{}}
$(0, 2, 3, 1)$, $(0, 3, 2, 1)$, $(2, 2, 2, 2)$, $(2, 2, 3, 1)$, $(2, 2, 3, 2)$, $(2, 2, 3, 3)$, $(2, 3, 2, 1)$, $(2, 3, 2, 2)$, 
\\
$(2, 3, 2, 3)$, $(4, 2, 2, 4)$, $(4, 2, 3, 3)$, $(4, 2, 3, 4)$, $(4, 2, 3, 5)$, $(4, 3, 2, 3)$, $(4, 3, 2, 4)$, $(4, 3, 2, 5)$ 
\end{tabular}
\\ \hline
$\leftindex_{\pm2}{\gamma}\leftindex_{\mp2}{\gamma}$ &
\begin{tabular}[c]{@{}l@{}}
$(0, 2, 2, 1)$, $(0, 2, 3, 1)$, $(0, 3, 2, 1)$, $(2, 2, 2, 1)$, $(2, 2, 2, 3)$, $(2, 2, 3, 1)$, $(2, 2, 3, 2)$, $(2, 2, 3, 3)$, 
\\
$(2, 3, 2, 1)$, $(2, 3, 2, 2)$, $(2, 3, 2, 3)$, $(4, 2, 2, 3)$, $(4, 2, 3, 3)$, $(4, 2, 3, 4)$, $(4, 2, 3, 5)$, $(4, 3, 2, 3)$, 
\\
$(4, 3, 2, 4)$, $(4, 3, 2, 5)$
\end{tabular}
\\ \hline
$\delta\leftindex_{\pm2}{\gamma}$ &
\begin{tabular}[c]{@{}l@{}}
$(0, 1, 2, 1)$, $(0, 2, 2, 1)$, $(0, 2, 3, 1)$, $(0, 3, 2, 1)$, $(1, 0, 2, 2)$, $(1, 1, 2, 1)$, $(1, 1, 2, 2)$, $(1, 1, 3, 2)$, 
\\
$(1, 2, 2, 0)$, $(1, 2, 2, 1)$, $(1, 2, 2, 2)$, $(2, 0, 2, 2)$, $(2, 0, 3, 3)$, $(2, 1, 2, 1)$, $(2, 1, 2, 2)$, $(2, 1, 2, 3)$, 
\\
$(2, 2, 2, 1)$, $(2, 2, 2, 2)$, $(2, 2, 2, 3)$, $(2, 2, 3, 1)$, $(2, 2, 3, 2)$, $(2, 2, 3, 3)$, $(2, 3, 2, 1)$, $(2, 3, 2, 2)$, 
\\
$(2, 3, 2, 3)$, $(3, 0, 2, 2)$, $(3, 1, 2, 2)$, $(3, 1, 2, 3)$, $(3, 1, 3, 2)$, $(3, 1, 3, 3)$, $(3, 1, 3, 4)$, $(3, 2, 2, 2)$, 
\\
$(3, 2, 2, 3)$, $(3, 2, 2, 4)$, $(4, 1, 2, 3)$, $(4, 2, 2, 3)$, $(4, 2, 2, 4)$, $(4, 2, 3, 3)$, $(4, 2, 3, 4)$, $(4, 2, 3, 5)$, 
\\
$(4, 3, 2, 3)$, $(4, 3, 2, 4)$, $(4, 3, 2, 5)$
\end{tabular}
\\ \hline
$u \leftindex_{\pm2}{\gamma}$ &
\begin{tabular}[c]{@{}l@{}}
$(1, 0, 2, 2)$, $(1, 1, 2, 1)$, $(1, 1, 2, 2)$, $(1, 1, 3, 2)$, $(1, 2, 2, 0)$, $(1, 2, 2, 1)$, $(1, 2, 2, 2)$, $(3, 0, 2, 2)$, 
\\
$(3, 1, 2, 2)$, $(3, 1, 2, 3)$, $(3, 1, 3, 2)$, $(3, 1, 3, 3)$, $(3, 1, 3, 4)$, $(3, 2, 2, 2)$, $(3, 2, 2, 3)$, $(3, 2, 2, 4)$ 
\end{tabular}
\\ \hline
\end{tabular}
\caption{Configurations of the nonvanishing TripoSH coefficients.}
\label{tab:triposh-coefficients}
\end{table}
We note that the dipolar modulation generates a larger number of nonvanishing \ac{TripoSH} coefficients than the 
$g_{LM}$-type quadrupolar modulation studied in ref.~\cite{Shiraishi:2023zda}. This difference arises from the distinct recoupling structure of the angular momenta, appearing in $c^{X}_j$, the \ac{CG} coefficients and the Wigner-$6j$ symbol. The corresponding configurations of the nonvanishing \ac{TripoSH} coefficients are listed in table~\ref{tab:triposh-coefficients}.
\subsection{The plane-parallel limit}\label{sec. The plane-parallel limit}
We move on to the \ac{PP} limit of eq.~\eqref{eq. Pi mod} by approximating $\hbmx_{1} \simeq \hbmx_{2} \simeq \hbmp$. In this limit, the correlation functions can be characterized by $\hbmx_{12}$ and $\hbmp$, so that a suitable basis becomes a set of the spin-weighted \ac{BipoSH}, defined by~\cite{Shiraishi:2023zda} 
\bae{\label{eq. BipoSH basis}
  \leftindex_{\lambda^\prime}{X}_{\ell \ell^\prime}^{LM}(\hbmx_{12}, \hbmp) 
 \coloneqq 
  \{\bm{Y}_{\ell}(\hbmx_{12}) \otimes 
  \leftindex_{\lambda^\prime}{\bm{Y}}_{\ell^\prime}(\hbmp)\}_{LM}
  = \sum_{m m^\prime} \calC_{\ell m \ell^\prime m^\prime}^{LM} Y_{\ell m}(\hbmx_{12}) \sY{\lambda^\prime}{\ell^\prime}{m^\prime}(\hbmp) ~.
}
The correlation function in the \ac{PP} limit is then expanded as
\bae{
  \xi_{\lambda_1 \lambda_2, \rm{PP}}^{X_1 X_2} (\bmx_{12},\hbmp) 
  = \sum_{\ell \ell^\prime L M} 
  \leftindex_{\lambda_1 \lambda_2}{\left(\hat{\xi}^{X_1 X_2}\right)}_{\ell\ell^\prime}^{LM} (x_{12}) 
  \leftindex_{\lambda_1+ \lambda_2}{X}_{\ell \ell^\prime}^{LM}(\hbmx_{12}, \hbmp) ~,
}
where the spin-weight index in the \ac{BipoSH} is given by the sum $\lambda_1 + \lambda_2$. This follows from the fact that, in the \ac{PP} limit, the spin-weighted \ac{TripoSH} are related to the spin-weighted \ac{BipoSH} as
\bae{
  \leftindex_{\lambda_1 \lambda_2}{\calX}_{\ell\ell_1 \ell_2 \ell^\prime}^{LM}(\hbmx_{12}, \hbmp, \hbmp) = \sqrt{\frac{(2\ell_1+1)(2\ell_2+1)}{4\pi (2\ell^\prime+1)}} (-1)^{\ell_1 + \ell_2 -\ell^\prime} \calC^{\ell^\prime \lambda_1+\lambda_2}_{\ell_1 \lambda_1 \ell_2 \lambda_2}
  \leftindex_{\lambda_1+\lambda_2}{X}_{\ell\ell^\prime}^{LM}(\hbmx_{12}, \hbmp) ~.
}
Analogously to the \ac{TripoSH} decomposition, we define the \ac{PP} limit of the position-dependent power spectrum as
\bae{\label{eq. PP limit TripoSH}
  P_{\lambda_1 \lambda_2, \rm{PP}}^{X_1 X_2} (\bm{k}, \hbmp)
  =
  \sum_{\ell \ell^\prime L M} 
  \leftindex_{\lambda_1 \lambda_2}{\left(\hat{\pi}^{X_1 X_2}\right)}_{\ell\ell^\prime}^{LM} (k) 
  \leftindex_{\lambda_1+ \lambda_2}{X}_{\ell \ell^\prime}^{LM}(\hbmk, \hbmp) ~.
}
The coefficient $\hat{\xi}$ can be obtained from $\hat{\pi}$ through the Hankel transform as
\bae{
\leftindex_{\lambda_1 \lambda_2}{\left(\hat{\xi}^{X_1 X_2}\right)}_{\ell\ell^\prime}^{LM} (x_{12}) = i^\ell \int_0^\infty \frac{k^2 \dd k}{2\pi^2} j_{\ell} (kx_{12}) \leftindex_{\lambda_1 \lambda_2}{\left(\hat{\pi}^{X_1 X_2}\right)}_{\ell\ell^\prime}^{LM} (k) ~,
}
where $\hat{\pi}$ is given by 
\bae{
\leftindex_{\lambda_1 \lambda_2}{\left(\hat{\pi}^{X_1 X_2}\right)}_{\ell \ell^\prime}^{LM} (k) = \int \dd\Omega_k \dd\Omega_{x_{\rm p}}
P_{\lambda_1 \lambda_2, \rm{PP}}^{X_1 X_2} (\bm{k}, \hbmp)
\leftindex_{\lambda_1+ \lambda_2}{X}_{\ell \ell^\prime}^{LM \ast}(\hbmk, \hbmp) ~.
}
Note that the coefficient $\hat{\pi}$ is related to $\hat{\Pi}$, introduced in eq.~\eqref{eq. hatPi}, by
\bae{
  \leftindex_{\lambda_1 \lambda_2}{\left(\hat{\pi}^{X_1 X_2}\right)}_{\ell\ell^\prime}^{LM} (k)  
   = 
   \sum_{\ell_1 \ell_2} (-1)^{\ell_1 + \ell_2 -\ell^\prime} 
  \sqrt{\frac{(2\ell_1+1)(2\ell_2+1)}{4\pi (2\ell^\prime+1)}} \calC^{\ell^\prime \lambda_1+\lambda_2}_{\ell_1 \lambda_1 \ell_2 \lambda_2}
  \leftindex_{\lambda_1 \lambda_2}{\left(\hat{\Pi}^{X_1 X_2}\right)}_{\ell\ell_1 \ell_2 \ell^\prime}^{LM} (k) ~.
}

Let us derive an explicit form of the coefficient $\hat{\pi}$ by writing it as the sum of two contributions:
\bae{\label{eq. small pi full}
  \leftindex_{\lambda_1 \lambda_2}{\left(\hat{\pi}^{X_1 X_2}\right)}_{\ell\ell^\prime}^{LM} (k) 
  = 
  \leftindex_{\lambda_1 \lambda_2}{\left(\pistd^{X_1 X_2}\right)}_{\ell \ell^\prime}^{LM} (k)
  +
  \leftindex_{\lambda_1 \lambda_2}{\left(\pimod^{X_1 X_2}\right)}_{\ell \ell^\prime}^{LM} (k) ~,
}
where $\pistd$ denotes the standard contribution in the statistically isotropic and homogeneous case, while $\pimod$ is the contribution from the dipolar modulation. We can easily show that these two contributions are given explicitly by
\bae{\label{eq. pi iso PP}
\leftindex_{\lambda_1 \lambda_2}{\left(\pistd^{X_1 X_2}\right)}_{\ell \ell^\prime}^{LM} (k) 
& \coloneqq 
\sum_{\ell_1 \ell_2} (-1)^{\lambda_1 + \lambda_2 + \ell_2 + \ell^\prime} \frac{4\pi}{\sqrt{2\ell^\prime+1}} \calC_{\ell_1 0 \ell_2 0}^{\ell 0}  \calC^{\ell^\prime \lambda_1+\lambda_2}_{\ell_1 \lambda_1 \ell_2 \lambda_2}
c^{X_1}_{\ell_1} c^{X_2}_{\ell_2} \delta^{\rm K}_{L,0} \delta^{\rm K}_{M,0} \delta^{\rm K}_{\ell, \ell^\prime} \bar{P}_{\rm m} (k) ~,
}
and
\bae{\label{eq. pi ani PP}
\leftindex_{\lambda_1 \lambda_2}{\left(\pimod^{X_1 X_2}\right)}_{\ell \ell^\prime}^{LM} (k)  
\coloneqq 
\sum_{\ell_1 \ell_2} (-1)^{\lambda_1 +\lambda_2}\left[\scrG^{(1,2)} + (-1)^{\ell} \scrG^{(2,1)}\right] A_{1 M} (k)  \delta^{\rm K}_{L,1}\bar{P}_{\rm m} (k) ~,
}
where we have introduced
\bae{
\scrG^{(1,2)} & \coloneqq
 (-1)^{\ell_1} \sum_{j}  \sqrt{\frac{4\pi (2\ell_1+1)^2}{2j+1}}
c_{j}^{X_1} c_{\ell_2}^{X_2}
  \calC^{j \lambda_1}_{1 0 \ell_1 \lambda_1}
  \calC^{\ell 0}_{j 0 \ell_2 0}
  \calC^{\ell^\prime \lambda_1+\lambda_2}_{\ell_1 \lambda_1 \ell_2 \lambda_2}
\begin{Bmatrix}
    \ell_1 & \ell_2 & \ell^\prime \\
    \ell & 1 & j
  \end{Bmatrix} ~,
\\
\scrG^{(2,1)} & \coloneqq \left. \scrG^{(1,2)}\right|_{(X_1, \ell_1, \lambda_1)\leftrightarrow (X_2, \ell_2, \lambda_2)} ~.
}
We note that the above expressions reproduce several known results in ref.~\cite{Shiraishi:2016wec} for $X_1=X_2=\delta$, up to the additional Doppler term, and the Legendre expansions for $\lambda_1 = \lambda_2 = 0$ presented in ref.~\cite{Minato:2025ozy}. See appendix~\ref{sec. check of PP limit} for the relation to the Legendre expansions.
\section{Numerical computation of the correlation functions}\label{sec. Numerical computation}
In this section, we numerically compute the correlation functions to demonstrate the need for the full-sky treatment. Motivated by the decaying behavior of the modulation suggested in ref.~\cite{Aiola:2015rqa}, we consider a phenomenological model with a global preferred direction $\bmd$:
\bae{
A_{1M} (k) = \frac{4\pi}{3} A \left(\frac{k}{\kc}\right)^{-1/2} Y^\ast_{1 M} (\hbmd)  ~,
}
where $A$ is the amplitude of the modulation and $\kc \coloneqq 0.005 ~\mathrm{Mpc}^{-1}$ is the pivot scale. Using the \ac{CMB} data, the amplitude is constrained to be $A = 0.063^{+ 0.028}_{- 0.030}$~\cite{Aiola:2015rqa}. In this paper, however, we treat $A$ as a free parameter satisfying $|A| \leq 0.1 $ for illustrative purposes. Using the ansatz above, the sum over the projection $M$ in the dipolar modulation~\eqref{eq. density spherical expression} leads to
\bae{\label{eq. modulation dx}
\sum_{M} A_{1M}(k) Y_{1 M} (\hbmx) =  A \left(\frac{k}{\kc}\right)^{-1/2} \hbmd \cdot \hbmx ~.
}

To focus on the comparison between the full-sky treatment and the \ac{PP} approximation, we consider a configuration in which the two galaxies are located at the same redshift.\footnote{When performing data analysis, we should not restrict ourselves to galaxy pairs at the same redshift to fully exploit the available information.}
\begin{figure}
\centering
\includegraphics[width=0.6\linewidth]{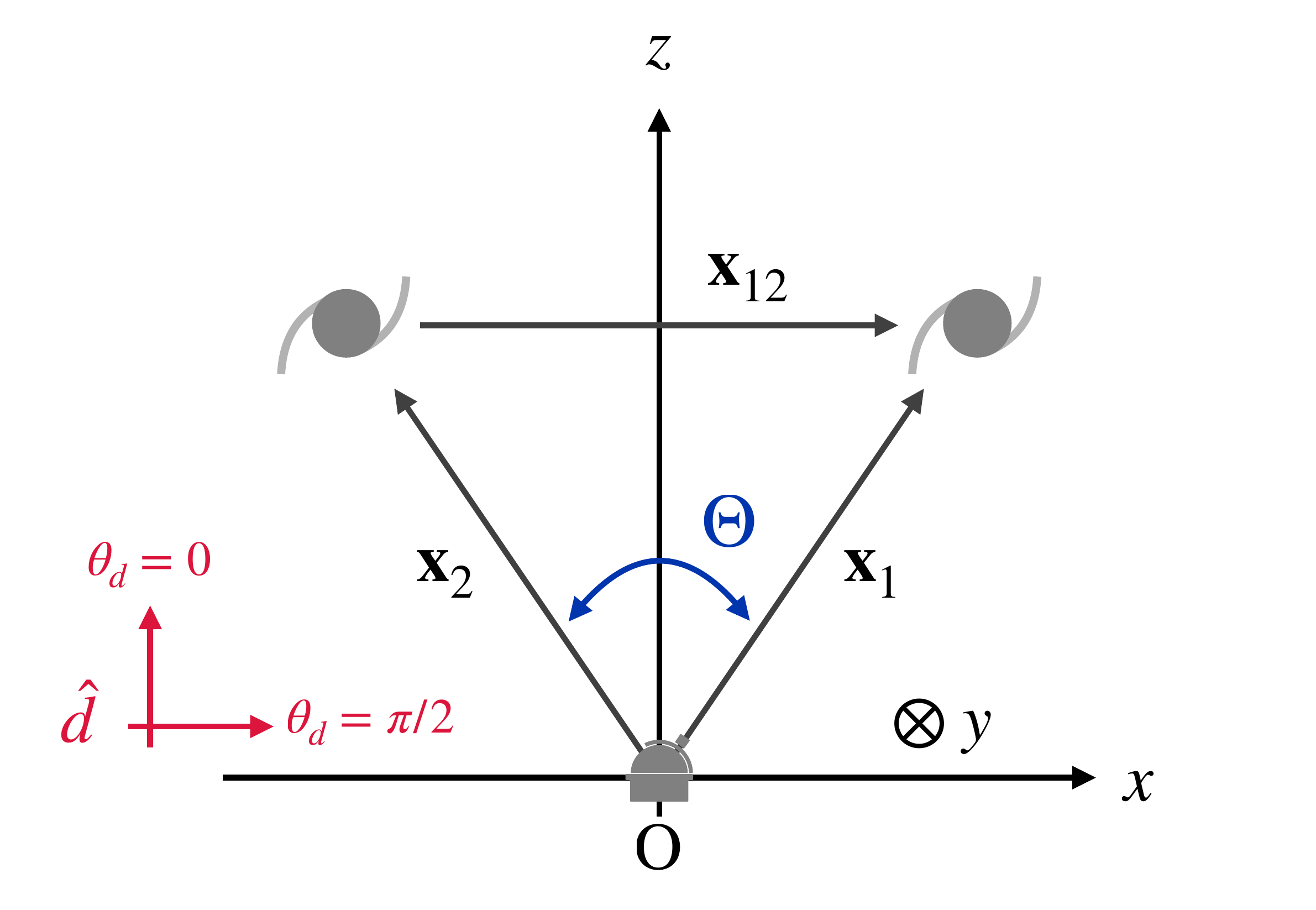}
\caption{Coordinate system adopted in the computation of the correlation functions.}
\label{fig: galaxies coordinate}
\end{figure}
We thus introduce the coordinate system where the three unit vectors $\hbmx_{1}$, $\hbmx_{2}$, and $\hbmx_{12}$ are set on the $xz$-plane as
\bae{
\hbmx_{1} = \left(\sin\frac{\Theta}{2}, 0, \cos\frac{\Theta}{2}\right) ~, 
\quad
\hbmx_{2} = \left(-\sin\frac{\Theta}{2}, 0, \cos\frac{\Theta}{2}\right) ~, 
\quad
\hbmx_{12} = \left(1, 0, 0\right) ~,
}
where $\Theta$ is the opening angle towards two target galaxies as shown in fig.~\ref{fig: galaxies coordinate}. 
The azimuthal angles are properly chosen such that the two \ac{LOS} directions are symmetrically oriented about the $z$-axis in the $xz$-plane. Note that, in the \ac{PP} limit, the \ac{LOS} direction is given by $\hbmp = (0,0,1)$. For the global preferred direction $\hbmd$, it can point to arbitrary directions. In the above-mentioned configuration, however, the contribution of the dipolar modulation to the correlation functions is independent of the azimuthal angle $\phi_d$, as can be seen from
\bae{\label{eq. modulation contribution}
\sum_{M} A_{1M}(k)\left[ Y_{1 M} (\hbmx_1) + Y_{1 M} (\hbmx_2) \right] \propto \hbmd \cdot \left(\hbmx_1 + \hbmx_2 \right) \propto \hbmd \cdot \hat{\bm{z}} = \cos\theta_d ~.
}
Thus, in the following discussion, we set $\phi_d = 0$, so that
\bae{
\hbmd=\left(\sin\theta_d,0,\cos\theta_d\right) ~.
}
For numerical calculations, we choose three values of the polar angle, $\theta_d \in \{0, \pi/4, \pi/2\}$.

\begin{figure}
    \centering
    \includegraphics[width=.9\linewidth]{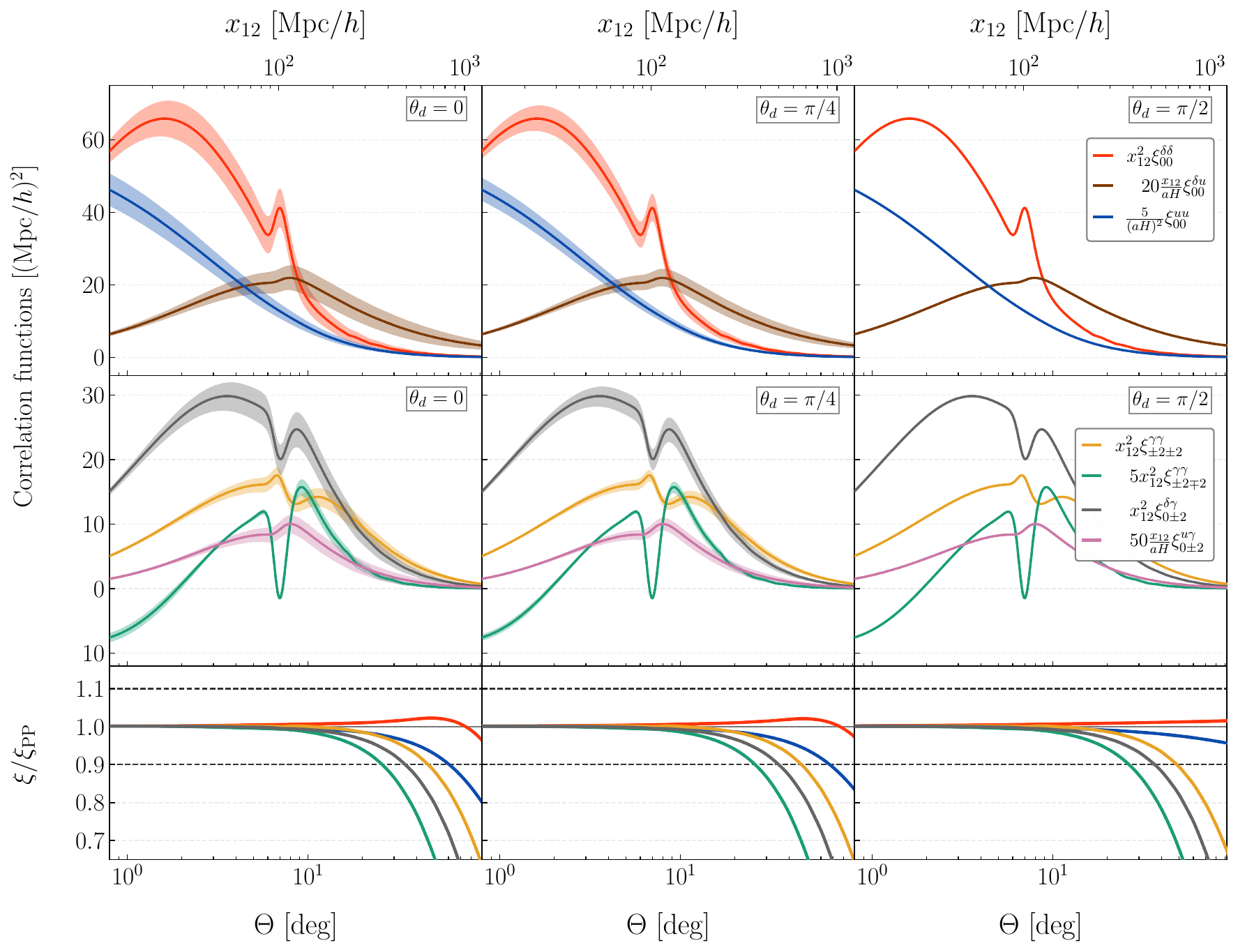}
    \caption{
    The correlation functions of the galaxy observables (the first and second rows) and the ratios between the exact results and their \ac{PP} limit (the bottom row) for $z_1=z_2=0.3$ and $b_g = b_{\rm K} =1$. We take $\alpha =2$ for the full correlation functions. In the first and second rows, solid lines correspond to the correlation functions with $A=0$ and the shaded bands indicate the ranges obtained by varying the dipolar modulation over $ - 0.1 < A < 0.1$. In the bottom row, we choose $A = 0.1$ as an example. 
    } 
    \label{fig: correlation functions}
\end{figure}

The correlation functions for the galaxy observables are shown in the first and second rows of fig.~\ref{fig: correlation functions}, where the matter power spectrum is computed using CAMB~\cite{Lewis:1999bs} with the Planck parameters~\cite{Planck:2018vyg}.
The baryon acoustic oscillation bump can be seen at $x_{12}\simeq 100 h^{-1}\mathrm{Mpc}$.
From left to right, we vary the value of $\theta_d$ and find that the contribution from the modulation is maximized when $\theta_d=0$ and decreases as $\theta_d$ approaches $\pi/2$.
This behavior is expected because, as shown in eq.~\eqref{eq. modulation contribution}, the amplitude of the contribution is proportional to $\cos \theta_d$ in the equal-redshift configuration. We emphasize, however, that the vanishing of the signal at $\theta_d=\pi/2$ is a consequence of the equal-redshift configuration. It is not generically true for the correlations between galaxies at different redshifts. We note that this overall $\theta_d$-dependence is peculiar to the dipolar position-dependent modulation. In models with the $g_{LM}$-type quadrupolar modulation, terms proportional to $\hbmd\cdot \hbmk$ lead to a non-trivial dependence on $\theta_d$, as illustrated in ref.~\cite{Shiraishi:2023zda}.

\begin{figure}
    \centering
    \includegraphics[width=.6\linewidth]{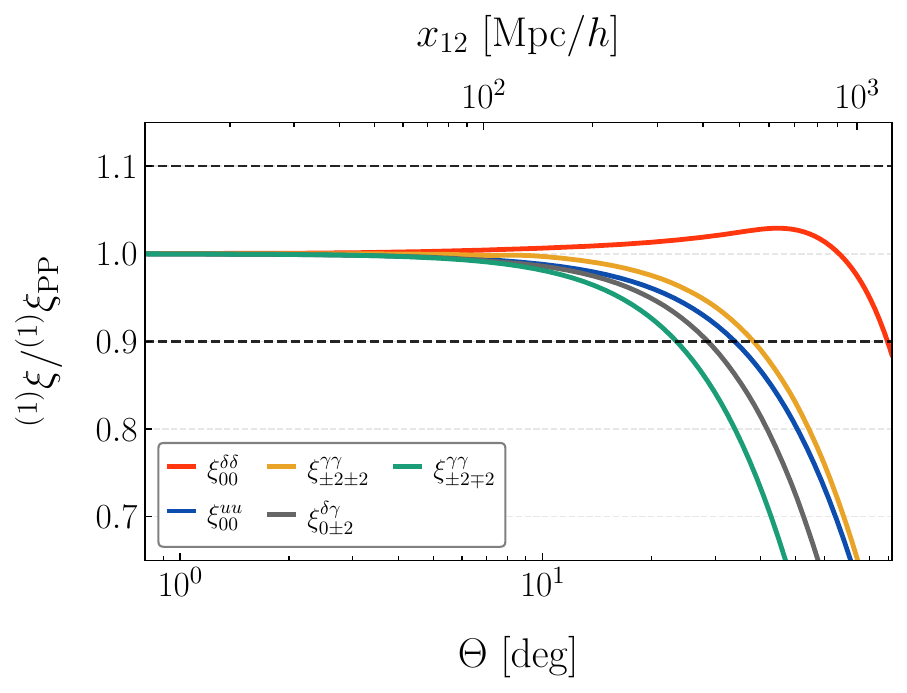}
    \caption{
    The ratios of the modulation contribution with and without the \ac{PP} approximation for $z_1=z_2=0.3$ and $b_g = b_{\rm K} =1$. We take $\alpha =2$ for the full correlation functions.
    } 
    \label{fig: correlation functions mod}
\end{figure}

Let us finally assess the importance of the full-sky treatment. In the bottom row of fig.~\ref{fig: correlation functions}, we show the ratios of the correlation functions computed with and without the \ac{PP} approximation, $\xi/\xi_{\rm PP}$, assuming $A = 0.1$. There, we omit the $\delta u$ and $u \gamma$ correlations because they vanish in the equal-redshift \ac{PP} configuration (see refs.~\cite{Burkey:2003rk,Howlett:2016urc} for the $\delta u$ case and ref.~\cite{Okumura:2019ned} for the $u \gamma$ case). We note, however, that they become non-vanishing beyond the \ac{PP} approximation~\cite{Castorina:2019hyr,Shiraishi:2020vvj,Shiraishi:2023zda}. The right panel corresponds to the case where the modulation contribution is absent due to the specific configuration with $\theta_d = \pi/2$. 
In that case, the full $\delta\delta$ correlation remains close to the \ac{PP} result. Similarly, one can see that the \ac{PP} approximation remains valid up to $\Theta \simeq \ang{90}$ within $10\%$ deviation for the $uu$ correlation~\cite{Shiraishi:2021oau}. For other correlations, errors exceed $10\%$ at opening angles of several tens of degrees. See ref.~\cite{Shiraishi:2020vvj} for the $\delta \gamma$ and $\gamma \gamma$ correlations.
In the presence of the large dipolar modulation, the deviation from the \ac{PP} limit can be larger, which is clear in fig.~\ref{fig: correlation functions mod}.
There, we focus on the modulation contribution of the correlation functions, which we denote ${}^{(1)}\xi$, and plot their ratios with and without the \ac{PP} approximation.
In the case of the position-dependent dipolar modulation, the ratios are independent of both the amplitude $A$ and the preferred-direction angle $\theta_d$, because the dependence on these quantities can be factored out of ${}^{(1)}\xi$ and cancels in the ratios.
This is clearly distinct from the $g_{LM}$-type modulation~\cite{Shiraishi:2023zda}, for which the ratios depend on $\theta_d$. Also, the ratios for the position-dependent modulation behave differently from those for the Alcock--Paczynski effect~\cite{Shiraishi:2021oau}. These differences reflect the distinct origins of the effects on the correlation functions. Despite these differences, all three cases exhibit a common trend: the \ac{PP} approximation becomes progressively less accurate for $\Theta \gtrsim \ang{30}$, making the wide-angle treatment increasingly important.
\section{Conclusions}\label{sec. Conclusions}
In this paper, we have provided a theoretical foundation for testing the dipolar modulation with three-dimensional two-point correlation functions in future wide-angle \ac{LSS} surveys. To this end, we utilize the spin-weighted \ac{TripoSH} decomposition as an efficient computational framework to treat both spin-weight $0$ observables, such as the galaxy density and peculiar velocity, and the spin-weight $\pm2$ ellipticity. To investigate the impact of the wide-angle effect, we have focused on the correlation functions for pairs of target galaxies located at the same redshift. In this restricted setup, the contribution from the modulation is controlled by the directional cosine between the pair-center direction and the preferred direction, as can be seen in eq.~\eqref{eq. modulation contribution}. Thus, the modulation effect is maximal when these two directions are parallel, and vanishes when they are perpendicular (see the first and second rows of fig.~\ref{fig: correlation functions}). We have also assessed the need for the wide-angle treatment by comparing the correlation functions with and without the \ac{PP} approximation. As can be seen in the bottom row of fig.~\ref{fig: correlation functions} and fig.~\ref{fig: correlation functions mod}, the \ac{PP} approximation becomes progressively unreliable for the opening angle $\Theta \gtrsim\ang{30}$. Our results therefore imply the importance of using the full-sky formalism when testing the dipolar modulation in ongoing and future \ac{LSS} surveys, such as DESI~\cite{DESI:2016fyo}, Euclid~\cite{EUCLID:2011zbd}, and SPHEREx~\cite{SPHEREx:2014bgr}.

This work can be regarded as a significant theoretical extension of the previous study on the galaxy density auto correlation alone~\cite{Shiraishi:2016wec}. Although we have focused on the dipolar modulation with $L=1$, which is directly motivated by the observed hemispherical power asymmetry in the \ac{CMB}, it would be natural to expect that the matter density field can also be modulated by higher multipoles with $L>1$. Such extensions are expected to differ from the dipolar case, both observationally and theoretically. We therefore leave a detailed analysis to a forthcoming paper~\cite{SMOO2}.
\section*{Acknowledgments}
TO acknowledges support from the Taiwan National Science and Technology Council under Grants 
Nos. NSTC 112-2112-M-001-034-,
NSTC 113-2112-M-001-011- and
NSTC 114-2112-M-001-004-, and the Academia Sinica Investigator Project Grant No. AS-IV-114-M03 for the period of 2025-2029. 
IO acknowledges support from JST FOREST Program under Contract No. JPMJFR222Y.
This work was supported in part by JSPS KAKENHI Grant Number JP23K03390 (MS).

\appendix
\section{Definitions and identities}\label{sec. mathematics}
\subsection{The spin-weighted spherical harmonics}\label{sec. The spin-weighted spherical harmonics}
Denoting $D^J_{M, M^\prime} (\alpha, \beta, \gamma)$ as the Wigner $D$-functions with $\{\alpha,\beta,\gamma\}$ being the Euler angles, the spin-weighted spherical harmonics are defined by~\cite{Gair:2015hra,Okamoto:2002ik,Shiraishi:2010sm}
\bae{
 \leftindex_{s}{Y}_{\ell m}(\theta, \phi) = (-1)^m \sqrt{\frac{2\ell+1}{4\pi}} D^\ell_{-m, s} (\phi, \theta, 0) ~.
}
The Wigner $D$-functions are given by the Wigner’s small $d$-function as~\cite{Varshalovich:1988ifq}
\bae{
D^J_{M, M^\prime} (\alpha, \beta, \gamma) = \ee^{-iM \alpha} d^J_{M, M^\prime} (\beta) \ee^{-iM^\prime \gamma} ~,
}
where the $d$-function may be represented by trigonometric functions as
\multi{
d^J_{M, M^\prime} (\beta) = (-1)^{J-M^\prime}\sqrt{(J+M)!(J-M)!(J+M^\prime)!(J-M^\prime)!} 
\\
\times \sum_{k} (-1)^k \frac{\sin^{2J}\left(\beta/2\right) \cot^{M+M^\prime + 2k}\left(\beta/2\right)}{k! (J-M-k)!(J-M^\prime-k)!(M + M^\prime + k)!} ~,
}
with $k$ running over all integer values for which the factorial arguments are non-negative. With this, we obtain the series representation of the spin-weighted spherical harmonics as~\cite{Gair:2014rwa}
\multi{
  \leftindex_{s}{Y}_{\ell m}(\theta, \phi) = (-1)^m \sqrt{\frac{(\ell +m)!(\ell-m)!(2\ell+1)}{(\ell+s)!(\ell-s)! 4\pi}} \sin^{2\ell}\left(\frac{\theta}{2}\right) 
  \\
  \times \sum_{k=0}^{\ell-s} \binom{\ell-s}{k} \binom{\ell+s}{k+s-m} (-1)^{\ell-k-s} \ee^{im\phi} \cot^{2k+s-m}\left(\frac{\theta}{2}\right) ~.
}
In the following, we summarize some properties of the spin-weighted spherical harmonics.

\noindent
Complex conjugation:
\bae{
  \leftindex_{s}{Y}^\ast_{\ell m} (\hbmx) = (-1)^{s+m} \leftindex_{-s}{Y}_{\ell -m} (\hbmx) ~.
}
Parity transformation:
\bae{
\leftindex_{s}{Y}_{\ell m} (-\hbmx) = (-1)^\ell \leftindex_{-s}{Y}_{\ell m} (\hbmx) ~.
}
Addition theorem:
\multi{\label{eq. the addition theorem}
  \leftindex_{s_1}{Y}_{\ell_1 m_1} (\hbmx) \leftindex_{s_2}{Y}_{\ell_2 m_2} (\hbmx) 
  \\
  =\sum_{s_3 \ell_3 m_3} (-1)^{-s_3 + m_3}
  \sqrt{\frac{(2\ell_1+1)(2\ell_2+1)}{4\pi(2\ell_3+1)}}\calC^{\ell_3 s_3}_{\ell_1 -s_1 \ell_2 -s_2}\calC^{\ell_3 -m_3}_{\ell_1 m_1 \ell_2 m_2} \leftindex_{s_3}{Y}^\ast_{\ell_3 m_3} (\hbmx) ~.
}
Orthonormality:
\bae{
\int \dd\Omega_x \leftindex_{s}{Y}_{\ell_1 m_1} (\hbmx)\leftindex_{s}{Y^\ast}_{\ell_2 m_2} (\hbmx) = \delta^{\rm K}_{\ell_1, \ell_2} \delta^{\rm K}_{m_1, m_2} ~.
}
\subsection{The Clebsch--Gordan coefficients}
We also summarize some properties of the \ac{CG} coefficients.

\noindent
Symmetry properties:
\bae{
\calC^{\ell_3 - m_3}_{\ell_1 - m_1 \ell_2 - m_2} & = (-1)^{-\ell_1 - \ell_2 + \ell_3} \calC^{\ell_3 m_3}_{\ell_1 m_1 \ell_2 m_2} ~,
\\
\calC^{\ell_3 m_3}_{\ell_1 m_1 \ell_2 m_2} & = (-1)^{\ell_1 + \ell_2 - \ell_3} \calC^{\ell_3 m_3}_{\ell_2 m_2 \ell_1 m_1} ~, 
\\
\calC^{\ell_3 m_3}_{\ell_1 m_1 \ell_2 - m_2} & = (-1)^{-\ell_2 + m_2} \sqrt{\frac{2\ell_3+1}{2\ell_1+1}}\calC^{\ell_1 m_1}_{\ell_2 m_2 \ell_3 m_3} ~,
\\
\calC^{\ell_3 m_3}_{\ell_1 -m_1 \ell_2 m_2} & = (-1)^{- \ell_1 - m_1} \sqrt{\frac{2\ell_3+1}{2\ell_2 + 1}}\calC^{\ell_2 m_2}_{\ell_3 m_3 \ell_1 m_1} ~.
}
The \ac{CG} coefficients with two of indices being zero:
\bae{
\calC_{ab00}^{cd} = \delta^{\rm K}_{a, c}\delta^{\rm K}_{b, d} ~.
}
Summation formulae involving \ac{CG} coefficients:
\bae{
  \sum_m (-1)^{-m} \calC^{L0}_{\ell m \ell -m} & = (-1)^{-\ell} \sqrt{2\ell+1} \delta^{\rm K}_{L, 0} ~,
\\
  \sum_{m_1 m_2} \calC^{\ell m}_{\ell_1 m_1 \ell_2 m_2}\calC^{\ell^\prime m^\prime}_{\ell_1 m_1 \ell_2 m_2} & = \delta^{\rm K}_{\ell, \ell^\prime} \delta^{\rm K}_{m, m^\prime} ~,
  \\
  \sum_{L M} \calC_{\ell_1 m_1 \ell_2 m_2}^{L M}\calC_{\ell_1 m_1^\prime \ell_2 m_2^\prime}^{L M} & = \delta^{\rm K}_{m_1, m_1^\prime} \delta^{\rm K}_{m_2, m_2^\prime} ~.
}
Summation formula to rewrite four \ac{CG} coefficients with the $6j$ symbol:
\multi{
  \sum 
  \calC^{jm}_{j_{12}m_{12} j_3 m_3}
  \calC^{j_{12}m_{12}}_{j_{1} m_{1} j_{2} m_{2}}
  \calC^{j^\prime m^\prime}_{j_{1} m_{1} j_{23} m_{23}}
  \calC^{j_{23} m_{23}}_{j_{2} m_{2} j_3 m_3}
  \\
  =
  \delta^{\rm K}_{j,j^\prime} \delta^{\rm K}_{m,m^\prime} (-1)^{j_{1} + j_{2} + j_3 + j}\sqrt{(2 j_{12} + 1)(2 j_{23}+1)}
  \left\{\begin{matrix}
    j_{1} & j_{2} & j_{12} \\
    j_3 & j & j_{23}
  \end{matrix} \right\} ~,
}
where the sum is performed over $m_1, m_2, m_3, m_{12}, m_{23}$.

\noindent
Summation involving products of the \ac{CG} coefficients and one $6j$ symbol:
\bae{\label{eq. 6j to sum CG}
\calC^{j_3 \epsilon}_{j_{12}\gamma j \phi}
\left\{\begin{matrix}
    j_{1} & j_{2} & j_{12} \\
    j_3 & j & j_{23}
  \end{matrix} \right\}
= (-1)^{- j_{1} - j_{2} - j_3 - j}\frac{1}{\sqrt{(2 j_{12}+1)(2 j_{23}+1)}} \sum_{\alpha\beta\delta} \calC_{j_{2}\beta j_{1}\alpha}^{j_{12}\gamma} \calC_{j_{2}\beta j_{23}\delta}^{j_3 \epsilon}
\calC_{j_{1}\alpha j \phi}^{j_{23}\delta} ~.
}
\section{Derivation of the unified expression for galaxy observables}\label{sec. derivation unified}
Let us first rewrite eqs.~\eqref{eq. Number density field} and \eqref{eq. peculiar velocity} in terms of the spherical harmonics. A unit vector $\hat{\bmk}$ can be expanded in the spherical harmonics with angular momentum $L=1$ as~\cite{Shiraishi:2010kd}
\bae{\label{eq. alpha vector}
  \hbmk = \sum_{m} \bm{\alpha}^m Y_{1m}(\hbmk) ~,
}
where the coefficient vectors $\bm{\alpha}^m$ are expressed in terms of the Kronecker delta $\delta^{\rm K}_{i,j}$ as
\bae{
 \bm{\alpha}^m = \sqrt{\frac{2\pi}{3}}
  \begin{pmatrix}
    -m \left(\delta^{\rm K}_{m,1} + \delta^{\rm K}_{m,-1}\right)
    \\
    i \left(\delta^{\rm K}_{m,1} + \delta^{\rm K}_{m,-1}\right)
    \\
    \sqrt{2}\delta^{\rm K}_{m,0}
    \end{pmatrix} ~.
}
Using the relation
\bae{
  \bm{\alpha}^m \cdot \bm{\alpha}^M = \frac{4\pi}{3}(-1)^m\delta^{\rm K}_{m, -M} ~,
}
one can easily show that eq.~\eqref{eq. Number density field} reduces to
\multi{\label{eq. Number density field Ylm}
\delta (\bm{x}) = 
\int \frac{\dd[3]{k}}{(2\pi)^3} 
\ee^{i \bm{k}\cdot\bm{x}}
\sum_{j\mu} \frac{4\pi}{2j+1}
  \left[\left(b_g+\frac{1}{3} f\right) \delta^{\rm K}_{j,0} - i\frac{\alpha (\bmx) f}{kx}\delta^{\rm K}_{j,1}+\frac{2}{3}f\delta^{\rm K}_{j,2}\right]
  \\
  \times Y_{j\mu}(\hbmk) Y^\ast_{j\mu} (\hbmx)\delta_{\rm m} (\bmk, \hbmx) ~,
}
and eq.~\eqref{eq. peculiar velocity} is expressed as
\bae{
\label{eq. peculiar velocity Ylm}
u (\bm{x}) & = \int\frac{\dd[3]{k}}{(2\pi)^3}\ee^{i \bm{k}\cdot\bm{x}} \sum_{j\mu} i \frac{a H f}{k} \frac{4\pi}{3}\delta^{\rm K}_{j,1} Y_{j\mu}(\hbmk) Y^\ast_{j\mu} (\hbmx)
\delta_{\rm m} (\bm{k}, \hbmx) ~.
}

Let us turn to the ellipticity~\eqref{eq. spin-weight ellipticity}. A function with nonzero spin-weight can be expanded in terms of the spin-weighted spherical harmonics $\leftindex_{\lambda}{Y}_{\ell m}$ with $\lambda$ being a spin-weight. See Appendix~\ref{sec. The spin-weighted spherical harmonics} for our conventions. In ref.~\cite{Shiraishi:2010kd}, it is shown that we can expand the basis vectors as
\bae{
  \bm{m}_{\pm} (\hbmx) = \pm \sum_{m} \bm{\alpha}^m \leftindex_{\mp 1}{Y}_{1 m} (\hbmx) ~,
}
where the vector $\bm{\alpha}^m$ is defined by eq.~\eqref{eq. alpha vector}. By using the addition theorem of the spin-weighted spherical harmonics~\eqref{eq. the addition theorem}, we obtain
\bae{
  m_{\mp}^i (\hbmx) m_{\mp}^j (\hbmx) \hatk_i \hatk_j = \frac{4\sqrt{6}\pi}{15} \sum_{\mu} Y_{2\mu} (\hbmk)\leftindex_{\mp 2}{Y}^\ast_{2\mu} (\hbmx) ~.
}
This leads to
\bae{\label{eq. spin-weight ellipticity Ylm}
  \leftindex_{\lambda}{\gamma} (\bm{x}) = \int\frac{\dd[3]{k}}{(2\pi)^3}\ee^{i \bm{k}\cdot\bm{x}}
  \sum_{j\mu} \frac{4\sqrt{6}\pi}{15} b_{\rm K} \delta^{\rm K}_{j,2} Y_{j\mu} (\hbmk)\leftindex_{-\lambda}{Y}^\ast_{j\mu} (\hbmx) \delta_{\rm m} (\bm{k}, \hbmx) ~,
}
with $\lambda = \pm2$.

\section{Derivation of plane-parallel limit with spin-weight $0$}\label{sec. check of PP limit}
We show that, for $\lambda_1=\lambda_2=0$, eq.~\eqref{eq. small pi full} recovers the result in ref.~\cite{Minato:2025ozy}. 
\subsection{The isotropic and homogeneous part}
Let us first focus on the standard isotropic and homogeneous part. The position-dependent power spectrum in the \ac{PP} limit, given in eq.~\eqref{eq. PP limit TripoSH}, can be written in terms of the Legendre polynomials $\calL_\ell (\mu)$ as
\bae{
P_{\rm{PP}}^{X_1 X_2} (\bm{k}, \hbmp) 
\supset
\sum_{\ell}\sum_{j_1 j_2} (-1)^{j_2} c^{X_1}_{j_1} c^{X_2}_{j_2} \left(\calC_{j_1 0 j_2 0}^{\ell 0}\right)^2 \calL_{\ell} (\mu) \bar{P}_{\rm m} (k) ~,
}
where $\mu = \hbmk\cdot \hbmp$ and 
\bae{
\calL_\ell (\mu) = \frac{4\pi}{2\ell+1}\sum_{m=-\ell}^{\ell} Y_{\ell m}(\hbmk) Y^\ast_{\ell m}(\hbmp) ~.
}
Here, we use the addition theorem~\eqref{eq. the addition theorem}. This defines the Legendre coefficients as 
\bae{\label{eq. the Legendre coefficients}
P^{X_1 X_2}_\ell (k) \coloneqq \sum_{j_1 j_2} (-1)^{j_2} c^{X_1}_{j_1} c^{X_2}_{j_2} \left(\calC_{j_1 0 j_2 0}^{\ell 0}\right)^2 \bar{P}_{\rm m} (k) ~,
}
with which eq.~\eqref{eq. pi iso PP} with $\lambda_1=\lambda_2=0$ is simplified to
\bae{
{\left(\pistd^{X_1 X_2}\right)}_{\ell \ell^\prime}^{LM} (k) 
=
\frac{4\pi}{2\ell+1}
\left\{\begin{pmatrix}
    \ell & \ell^\prime & 0 \\
    0 & 0 & 0
  \end{pmatrix}\right\}^{-1} 
  \delta^{\rm K}_{L,0} \delta^{\rm K}_{M,0} \delta^{\rm K}_{\ell, \ell^\prime} P^{X_1 X_2}_\ell (k) ~.
}
This agrees with the result shown in ref.~\cite{Minato:2025ozy}.
\subsection{The modulation part}
Recall from eq.~\eqref{eq. pi ani PP} that $\pimod$ is given by
\bae{
\leftindex_{\lambda_1 \lambda_2}{\left(\pimod^{X_1 X_2}\right)}_{\ell \ell^\prime}^{LM} (k)  
=
\sum_{\ell_1 \ell_2} (-1)^{\lambda_1 +\lambda_2} \left[\scrG^{(1,2)} + (-1)^{\ell} \scrG^{(2,1)}\right] A_{1 M} (k)  \delta^{\rm K}_{L,1}\bar{P}_{\rm m} (k) ~,
}
with
\bae{
\scrG^{(1,2)} & =
 (-1)^{\ell_1} \sum_j \sqrt{\frac{4\pi (2\ell_1+1)^2}{2j+1}}
c_{j}^{X_1} c_{\ell_2}^{X_2}
\calC^{j \lambda_1}_{1 0 \ell_1 \lambda_1}
  \calC^{\ell 0}_{j 0 \ell_2 0}
  \calC^{\ell^\prime \lambda_1+\lambda_2}_{\ell_1 \lambda_1 \ell_2 \lambda_2}
\begin{Bmatrix}
    \ell_1 & \ell_2 & \ell^\prime \\
    \ell & 1 & j
  \end{Bmatrix} ~,
\\
\scrG^{(2,1)} & = \left. \scrG^{(1,2)}\right|_{(X_1, \ell_1, \lambda_1)\leftrightarrow (X_2, \ell_2, \lambda_2)} ~.
}
When $\lambda_1=\lambda_2=0$, eq.~\eqref{eq. 6j to sum CG} can be written as
\bae{
\calC_{b 0 a 0}^{\ell^\prime 0}
\begin{Bmatrix}
    b & a & \ell^\prime \\
    \ell & 1 & j
  \end{Bmatrix}
  = (-1)^{- 1 - j -\ell^\prime - a} \frac{1}{\sqrt{(2b+1)(2\ell+1)}} \sum_{\sigma} \calC^{b 0}_{1 - \sigma j \sigma} \calC_{1 - \sigma \ell \sigma}^{\ell^\prime 0} \calC_{j \sigma a 0}^{\ell \sigma} ~,
  }
with which we have
\bae{
\sum_{\ell_1 \ell_2}\scrG^{(1,2)} & = \sum_{\ell_2 j} (-1)^{\ell - \ell_2}\sqrt{\frac{4\pi}{2\ell + 1}} c^{X_1}_{j} c^{X_2}_{\ell_2} \calC_{10\ell 0}^{\ell^\prime 0} \left(\calC_{j 0\ell_2 0}^{\ell 0}\right)^2 ~,
\\
\sum_{\ell_1 \ell_2}\scrG^{(2,1)} & = \sum_{\ell_1 j} (-1)^{\ell - \ell_1}\sqrt{\frac{4\pi}{2\ell + 1}} c^{X_2}_{j} c^{X_1}_{\ell_1} \calC_{10\ell 0}^{\ell^\prime 0} \left(\calC_{j 0\ell_1 0}^{\ell 0}\right)^2 ~.
}
Using eq.~\eqref{eq. the Legendre coefficients}, we then finally obtain 
\bae{
{\left(\pimod^{X_1 X_2}\right)}_{\ell \ell^\prime}^{LM} (k) 
& = (-1)^\ell \sqrt{\frac{16\pi}{2\ell+1}} \calC_{1 0 \ell 0}^{\ell^\prime 0} A_{1 M} (k)
\delta^{\rm K}_{L,1} P^{X_1 X_2}_\ell (k) ~,
\\
& = - \sqrt{\frac{16\pi(2\ell^\prime +1)}{2\ell+1}}\left(\begin{matrix}
    \ell & \ell^\prime & 1 \\
    0 & 0 & 0
  \end{matrix} \right) A_{1M} (k)\delta^{\rm K}_{L, 1} P^{X_1 X_2}_\ell (k) ~,
}
which shows agreement with the result in ref.~\cite{Minato:2025ozy} after replacing $A_{1M}\to A_{1M} f_{\rm mod}(k)$.
\bibliographystyle{JCAP}
\bibliography{Bib}
\end{document}